# A spline interpolation based data reconstruction technique for estimation of strain time constant in ultrasound poroelastography

Md Tauhidul Islam, Raffaella Righetti*

## Abstract

Ultrasound poroelastography is a cost-effective non-invasive imaging technique, which is able to reconstruct several mechanical parameters of cancer and normal tissue such as Young's modulus, Poisson's ratio, interstitial permeability and vascular permeability. To estimate the permeabilities, estimation of the strain time constant (TC) is required, which is a challenging task because of non-linearity of the exponential strain curve and noise present in the experimental data. Moreover, noise in many strain frames becomes very high because of motion artifacts from the sonographer, animal/patient and/or the environment. Therefore, using these frames in computation of strain TC can lead to inaccurate estimates of the mechanical parameters. In this letter, we introduce a cubic spline based interpolation method, which uses only the good frames (frame of high SNR) to reconstruct the information of the bad frames (frames of low SNR) and estimate the strain TC. We prove with finite element simulation that the proposed reconstruction method can improve the estimation accuracy of the strain TC by 46% in comparison to the estimates from noisy data, and 37% in comparison to the estimates from Kalman filtered data at an SNR of 30 dB. Based on the high accuracy of the proposed method in estimating strain TC from poroelastography data, the proposed method can be preferred technique by the clinicians and researchers interested in non-invasive imaging of tissue mechanical parameters.

This work was supported in part by the U.S. Department of Defense under grant W81XWH-18-1-0544 (BC171600).

Md Tauhidul Islam and *R. Righetti are with the Department of Electrical and Computer Engineering, Texas A&M University, College Station, TX 77843 USA (e-mail: righetti@ece. tamu.edu).

2## I. Introduction

Ultrasound elastography is a popular non-invasive method for measuring the stiffness of soft tissue based on induced strain from a small applied compression [1], [2]. This method is used as a diagnostic tool in many diseases, where the tissue stiffness changes at the onset of the disease. Poroelastography is a branch of elastography technique, where the temporal mechanical behavior of tissue is imaged in terms of induced strain under sustained compression [3]–[5]. The temporal profiles of strains from poroelastography experiments are related to the fluid transport properties of the tissue: interstitial permeability and vascular permeability and steady state strains are linked to the linear elastic properties: Young's modulus and Poisson's ratio. The motivation behind poroelastography is that in many diseases as cancers, the fluid transport properties of the tissue change. The time constant of the induced strains in poroelastography is important as it carries information of the fluid transport properties of the tissue [6], [7].

In poroelastography, to estimate the temporal profiles of strains with reduced decorrelation noise, strains between successive frames are estimated and then summed cumulatively [8]. However, in experimental conditions, because of motion artifacts created from the movement of the sonographer, movement/breathing of the animal/patient, environmental causes, many strain frames can still be of very low SNR (even $< 0$ dB) [9]. As a result, the SNR of the estimated cumulative strains reduces significantly [10].

Generally, the curve-fitting techniques are used for estimating the strain TC. However, curve fitting based techniques have known limitations, especially if the underlying model requires fitting a nonlinear function (as it is the case in poroelastography) since very small changes in the data can result in significantly different estimated parameters [11]. Moreover, these methods are very sensitive to noise and can fail or reach to a totally wrong solution, if the data is of low SNR [12]. These methods also can reach local minima in place of global one in data of low SNR.

There are a number of methods proposed in the literature to estimate the strain TC. Levenberg-Marquardth (LM) method has been utilized by several groups including ours to estimate the strain TC [12]. However, this method is prone to error in noisy conditions. In [13]–[15], the authors used traditional curve peeling/stripping method at first and then fine-tuned the obtained estimates using a nonlinear Gauss-Newton fitting technique blended with the LM method. In [16], the authors estimated the strain TC and steady state values based on searching poles and zeros in the Laplace domain. However, all these methods are modestly robust in noise and may fail to provide reasonable estimates in noisy experimental conditions.

Recently, we proposed a curve-fitting method based on variable projection, which can estimate the strain TC from data of low SNR such as 30 dB [10]. However, in many practical scenarios, the strain data be can be of much lower quality ($< 0$ dB) and a method that can denoise the strain data for better estimation of the strain TC can be very useful. In this letter, we propose a cubic spline interpolation based method for denoising the strain data before using curve-fitting technique for estimating the strain TC, which considers only the frames of high SNR. As this method does not use the frames of low SNR, it maintains high SNR and provides strain TC images of high quality.

## II. Problem formulation

In poroelastography, the first RF frame is the pre-compression frame and successive RF frames are post-compression frames at different time points under sustained compression. The axial strain between successive frames at a single pixel as a function of time can be expressed as [8]

$$x(t) = -\frac{\gamma}{\tau}e^{-\frac{t}{\tau}}, \tag{1}$$

whete $\tau$ is the time constant of the exponential axial strain curve and $\gamma$ is a constant parameter, which differs based on material properties and experimental protocol [10].

Integrating eq. (1) (equivalent to cumulative sum of discrete data), we get a generalized equation for the axial strain in a poroelastic sample during creep compression [12], [17]

$$s(t) = \eta + \gamma e^{-\frac{t}{\tau}}, \tag{2}$$

where $s(t)$ is the axial strain temporal curve, $\eta$ is the value of $s$ at steady state (i.e., at $t = \infty$).

As in practical experiments, the data is acquired discretely, Eq. (1) can be written as

$$x[n] = -\frac{\gamma}{\tau}e^{-\frac{nT_s}{\tau}}, \quad n = 1, \ldots, N, \tag{3}$$

where $T_s$ is the sampling time and $N$ is the total number of time samples.

## III. Cubic spline based frame reconstruction

Let us assume that among the available $N$ data points, $N_g$ data points are of higher SNR and $N - N_g$ are of lower SNR. For $N_g$ available good data points $x_i$, $i = 1, \ldots, N_g$, the essential idea behind the cubic spline based data reconstruction is to fit a piecewise function of the form [18]

$$S(x) = \begin{cases} s_1(x) & \text{if } x_1 \leq x < x_2 \\ s_2(x) & \text{if } x_2 \leq x < x_3 \\ \vdots & \\ s_{N_g-1}(x) & \text{if } x_{N_g-1} \leq x < x_{N_g} \end{cases}$$

where $s_m$ is a third degree polynomial defined by

$$s_m(x) = a_m(x - x_m)^3 + b_m(x - x_m)^2 + c_m(x - x_m) + d_m,$$

$$m = 1, \ldots, N_g - 1. \tag{4}$$

The cubic spline needs to satisfy certain properties. First, the function $S(x)$ will interpolate all data points. Second, $S(x), S'(x)$ and $S''(x)$ are continuous in the interval $[x_1, x_{N_g}]$. Using these properties and second derivative be equal to zero at the endpoints ($x[1]$ and $x[N]$) (for natural spline), parameters for generating a unique cubic spline can be estimated [18]. The readers are referred to [18] for further knowledge on cubic spline based interpolation.

## IV. SIMULATIONS

A commercial finite element (FE) simulation software namely Abaqus, Dassault Systemes Simulia Corp., Providence, RI, USA was used to simulate a cylindrical poroelastic sample containing a spherical inclusion under boundary conditions mimicking those of ultrasound elastography experiments. An instantaneous uniaxial stress of 1 kPa was applied on the sample for creep experiment simulation. The total analysis was recorded for 150 s in each step of 0.5 s. The readers are referred to our previous work [19] for detail description of the procedure for poroelastic simulation of a cylindrical sample with spherical inclusion.

The sample specification is given in Table I. In all samples, the interstitial permeability is assumed to be $3.1 \times 10^{-14}$ $m^4 N^{-1} s^{-1}$ in inclusion and $6.4 \times 10^{-15}$ $m^4 N^{-1} s^{-1}$ in background region [20], [21]. The microfiltration coefficient is assumed to be $1.09 \times 10^{-6}$ (Pa s)$^{-1}$ in inclusion and $4.44 \times 10^{-7}$ (Pa s)$^{-1}$ in background region [20], [21]. The cylindrical sample has a radius of 2 cm and height of 4 cm. The inclusion has a radius of 0.75 cm.

To create noisy strain data mimicking the ultrasound strain data, Gaussian noise is added to the strain data at different SNR (30, 40 and 60 dB) [22]. The SNR of the bad strain frames is set to 0 dB. The temporal positions of the bad frames are selected randomly from a discrete uniform distribution of frame numbers in the range of 1 to $N$.

TABLE I: Properties of the simulated poroelastic samples. $E, \nu$ and $\tau$ denotes the Young's modulus, Poisson's ratio and strain TC. Subscripts $i$ and $b$ denotes the mechanical parameters corresponding to inclusion (tumor) and background (normal tissue) region.

| Sample name | $E_i$ (kPa) | $\nu_i$ | $\tau_i$ (s) | $E_b$ (kPa) | $\nu_b$ | $\tau_b$ |
|---|---|---|---|---|---|---|
| A | 49.17 | 0.45 | 4.66 | 32.78 | 0.47 | 11.42 |
| B | 97.02 | 0.45 | 2.36 | 32.78 | 0.47 | 11.42 |
| C | 63.90 | 0.47 | 2.26 | 32.78 | 0.49 | 3.08 |

The performance of the proposed method is compared with that of Kalman filter, which is normally used in reducing noise in elastographic strain data [23]. In the Kalman filter, the length of Kalman window was taken as 13 [23].

Eq. (2) is fit onto the noisy axial strain data and denoised strain data using the proposed method and Kalman filter by the curve fitting tool in Matlab, Mathworks Inc., Natick, MA, USA to estimate the strain TC [10]. A least squared error minimization technique based on 'Levenberg-Marquardte' method is used for curve-fitting.

A. Image quality analysis for simulated data

Quality of strain TC images was quantified using PRE (percent relative error). The PRE is defined as

$$\text{PRE} = \frac{\rho_e - \rho_t}{\rho_t} \times 100, \tag{5}$$

where $\rho_e$ is the mean estimated strain TC inside the sample and $\rho_t$ is the true parameter (from the ideal model shown in Table I).

V. RESULTS

A. Simulations

Fig. 1 shows the results of curve-fitting on the clean signal from FEM, noisy signal of 60 dB SNR, Kalman filtered data and spline interpolated data, when $75\%$ of the strain frames are of high SNR. From this figure, we see that the exponential curves fitted to the clean and spline interpolated data has the TC close to the ideal value (Table I). On the other hand, the exponential curves could not be fitted to the noisy and Kalman filtered data properly and the obtained estimates of TC from these data are far from the true value.

Figs. 2 shows the resulting strain TC images using noisy, Kalman filtered and spline interpolated data for data of 60 dB SNR for different numbers of good frames, i.e., 75%, 50% and 20% of total number of frames. We see that in most pixels, the obtained estimates of the strain TC in noisy and Kalman filtered data are far from true value and the images are overall very noisy. The TC images obtained using spline interpolated data has values closer to the true value. With decrement of number of good frames, the computed TC values for the spline interpolated data start to deviate from the true value as seen in Figs. 2(C3). The noise in the estimated strain TC from noisy and Kalman filtered data increases also as number of good frames decreases.

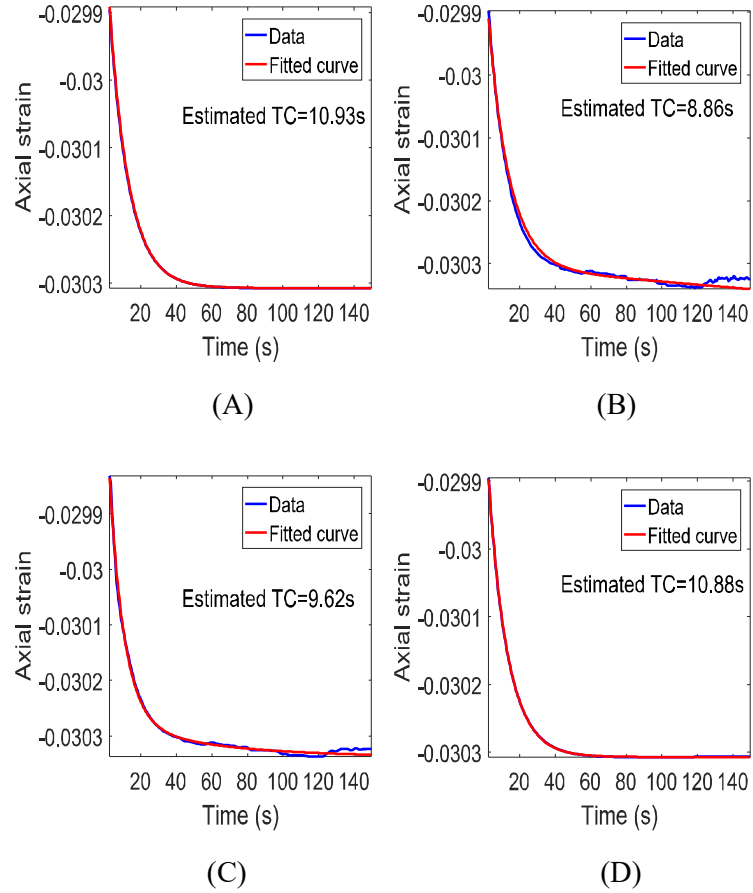

Fig. 1: Axial strain temporal data with fitted curves (A) clean data (B) noisy data at SNR of 60 dB (c) Kalman filtered data (D) spline interpolated data.

Similar scenario can be seen for the data of 30 dB SNR as shown in Fig. 3 for different numbers of good frames, i.e., 75%, 50% and 20% of total frame number. We see that even in case of 30 dB SNR, the strain TC estimates from spline interpolated data is almost noise free and have values close to the true value. Overall, in case of 30 dB SNR, the noise in estimated strain TC images from noisy, Kalman filtered and spline interpolated data increases when compared to the estimated strain TC images from data of 60 dB SNR.

The PRE in estimated strain TC images from noisy, Kalman filtered and spline interpolated data at 60 dB SNR for sample A are shown in Table II. From this table, we see that error in highest in case of noisy data and lowest in case of spline interpolated data for all considered cases. The error can be as high as 55.26% in case of noisy data at 30 dB SNR with 20% of total frame being of high SNR, where the PRE in estimated strain by the proposed method is 8.17% and by Kalman filter 45.11%. For the spline interpolated data, the error can be as low as 5.61% at 60 dB SNR when 75% of the frames are of high SNR, where the PRE in estimated TC from noisy data is 32.05% and from Kalman filtered data 28.31%. Overall, PRE reduces for all the methods, when number of high SNR frames increases.

The computed errors in samples B and C are shown in Table III and IV. Overall, the error increases at all SNRs in sample C and decreases in sample B when compared to sample A. This may be because of higher Young's modulus contrast in sample B and lower TC contrast in sample C.

TABLE II: PRE (%) in estimated TC at different SNR levels for different percentage of good frame (PGF) for noisy, Kalman filtered and spline reconstructed data in sample A

| PGF (%) | 20 | | | 50 | | | 75 | | |
|---|---|---|---|---|---|---|---|---|---|
| SNR (dB) | 30 | 40 | 60 | 30 | 40 | 60 | 30 | 40 | 60 |
| Noisy | 55.26 | 54.38 | 54.26 | 43.87 | 40.78 | 40.46 | 34.39 | 33.85 | 32.05 |
| Kalman | 45.11 | 44.42 | 44.68 | 39.08 | 38.28 | 37.99 | 33.66 | 33.52 | 28.31 |
| Spline | 8.17 | 8.11 | 8.07 | 7.09 | 6.28 | 6.27 | 6.69 | 5.91 | 5.61 |

TABLE III: PRE (%) in estimated TC at different SNR levels for different percentage of good frame (PGF) for noisy, Kalman filtered and spline reconstructed data in sample B

| PGF (%) | 20 | | | 50 | | | 75 | | |
|---|---|---|---|---|---|---|---|---|---|
| SNR (dB) | 30 | 40 | 60 | 30 | 40 | 60 | 30 | 40 | 60 |
| Noisy | 56.48 | 54.93 | 55.39 | 58.13 | 51.84 | 55.05 | 46.29 | 41.18 | 40.94 |
| Kalman | 47.04 | 48.03 | 45.77 | 50.04 | 49.10 | 47.29 | 45.39 | 39.21 | 37.84 |
| Spline | 9.64 | 9.43 | 9.11 | 8.83 | 7.98 | 7.53 | 7.18 | 7.02 | 6.53 |

## VI. Discussion

In this letter, we propose a novel technique based on cubic spline interpolation for estimation of strain TC from axial strains in a poroelastography experiment. The proposed technique has been found to be accurate and robust to the noise at different SNR levels and for different number of good frames.

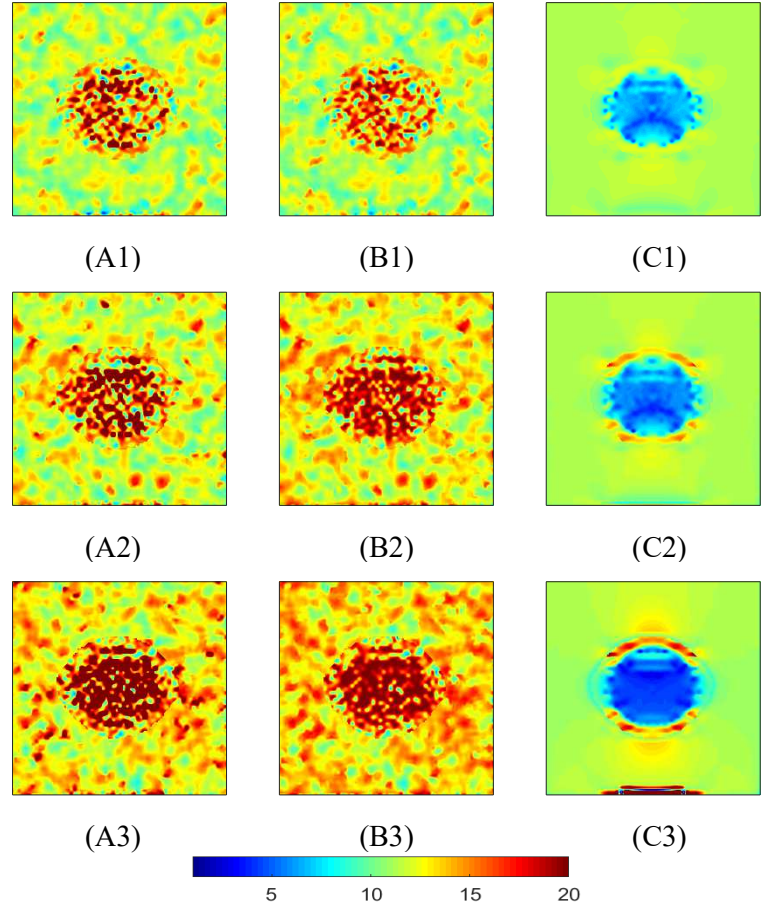

Fig. 2: Estimated strain TC images (in second) at 60 dB SNR for (1) 75% (2) 50% (3) 20% of total strain frames being of high SNR from (A) noisy strain data (B) Kalman filtered data and (C) spline interpolated data.

The computation time for the proposed technique is similar to that required for noisy strain TC estimation, which is much less than that required for Kalman filtered data. For a strain TC image of $128 \times 128$ pixel$^2$, the computation times required for noisy, spline interpolated and Kalman filtered data are 451.6 s, 455.72 s and 605.04 s, respectively. Further improvement in computation of the proposed method may be achieved by GPU (graphics processing unit) implementation and left for future work.

We have chosen the cubic spline interpolation over other interpolation techniques for re-constructing the strain data at frames of lower SNR for several reasons. First of all, spline interpolation is better than polynomial interpolation because the interpolation error can be made small using low order of spline in comparison to the same order of polynomials [24]. Second, the problem of Runge's phenomenon does not arise in spline interpolation, in which oscillation can occur between points when high degree polynomials is used for interpolation [24].

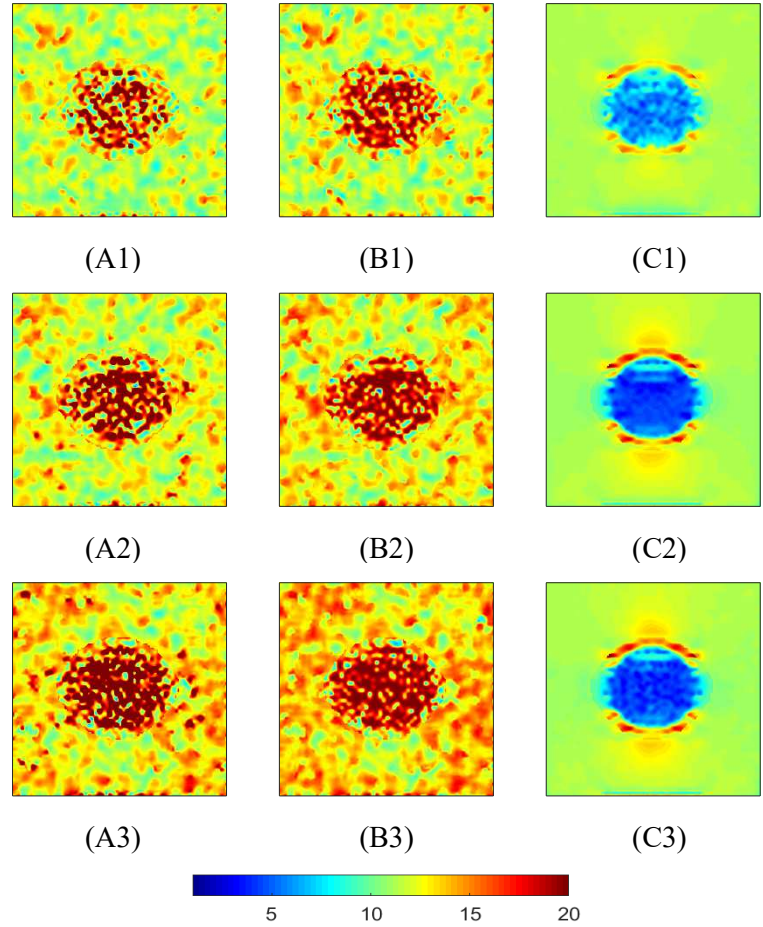

Fig. 3: Estimated strain TC images (in second) at 30 dB SNR for (1) 75% (2) 50% (3) 20% of total strain frames being of high SNR from (A) noisy strain data (B) Kalman filtered data and (C) spline interpolated data.

## VII. Conclusion

In this paper, we propose a cubic spline interpolation based data reconstruction technique for estimation of TC from axial strain data in ultrasound poroelastography. The proposed method is found to be highly accurate and robust to noise. Based on the importance of strain TC in estimation of mechanical parameters from poroelastography data, the proposed methodology is expected to keep important role in quantitative cancer imaging.

TABLE IV: PRE (%) in estimated TC at different SNR levels for different percentage of good frame (PGF) for noisy, Kalman filtered and spline reconstructed data in sample C

| PGF (%) | 20 | | | 50 | | | 75 | | |
|---|---|---|---|---|---|---|---|---|---|
| SNR (dB) | 30 | 40 | 60 | 30 | 40 | 60 | 30 | 40 | 60 |
| Noisy | 49.47 | 41.95 | 38.98 | 35.32 | 34.90 | 34.77 | 32.28 | 32.07 | 26.39 |
| Kalman | 41.59 | 37.11 | 32.13 | 31.98 | 34.50 | 29.75 | 29.34 | 28.52 | 25.22 |
| Spline | 7.09 | 7.74 | 7.13 | 6.33 | 5.19 | 4.93 | 4.69 | 3.98 | 4.25 |